\documentclass[aps,prl,twocolumn,superscriptaddress,nobibnotes,nodoi,noeprint,longbibliography]{revtex4-2}
\usepackage{amsmath,amssymb,physics,bm,siunitx}
\usepackage{xcolor,graphicx}
\usepackage{soul}
\usepackage[colorlinks=true,citecolor=blue,urlcolor=blue,linkcolor=black]{hyperref}

\newcommand{\change}[1]{{#1}}

\newcommand{\rms}{\rm\scriptscriptstyle}

\begin{document}
\title{Hydrodynamic interactions can induce jamming in flow-driven systems}
\author{Eric Cereceda-L\'opez}
\affiliation{Departament de F\'{i}sica de la Mat\`{e}ria Condensada, Universitat de Barcelona, 08028, Spain}
\affiliation{Institut de Nanoci\`{e}ncia i Nanotecnologia, Universitat de Barcelona (IN2UB), 08028, Barcelona, Spain}
\author{Dominik Lips}
\affiliation{Universit\"{a}t Osnabr\"{u}ck, Fachbereich Physik, Barbarastra{\ss}e 7, D-49076 Osnabr\"uck, Germany}
\author{Antonio Ortiz-Ambriz}
\email{aortiza@fmc.ub.edu}
\affiliation{Departament de F\'{i}sica de la Mat\`{e}ria Condensada, Universitat de Barcelona, 08028, Spain}
\affiliation{Institut de Nanoci\`{e}ncia i Nanotecnologia, Universitat de Barcelona (IN2UB), 08028, Barcelona, Spain}
\affiliation{University of Barcelona Institute of Complex Systems (UBICS), 08028, Barcelona, Spain}
\author{Artem Ryabov}
\email{artem.ryabov@mff.cuni.cz}
\affiliation{Charles University, Faculty of Mathematics and Physics, Department of Macromolecular Physics, V Hole\v{s}ovi\v{c}k\'{a}ch 2, 
CZ-18000 Praha 8, Czech Republic}
\author{Philipp Maass}
\email{maass@uos.de}
\affiliation{Universit\"{a}t Osnabr\"{u}ck, Fachbereich Physik, Barbarastra{\ss}e 7, D-49076 Osnabr\"uck, Germany}
\author{Pietro Tierno}
\affiliation{Departament de F\'{i}sica de la Mat\`{e}ria Condensada, Universitat de Barcelona, 08028, Spain}
\affiliation{Institut de Nanoci\`{e}ncia i Nanotecnologia, Universitat de Barcelona (IN2UB), 08028, Barcelona, Spain}
\affiliation{University of Barcelona Institute of Complex Systems (UBICS), 08028, Barcelona, Spain}

\date{\today} 

\begin{abstract}
Hydrodynamic interactions between fluid-dispersed particles are ubiquitous in soft matter and biological systems and they give rise to intriguing collective phenomena. 
While it was reported that these interactions can facilitate force-driven particle motion over energetic barriers, here we show the opposite effect in a flow-driven system, i.e. that hydrodynamic interactions hinder transport across barriers. 
We demonstrate this result by combining experiments and theory. In the experiments, we drive colloidal particles using rotating optical traps, thus creating a vortex flow in the corotating reference frame. 
We observe a jamming-like decrease of particle currents with density for large barriers between traps. The theoretical model shows that this jamming arises from hydrodynamic interactions between the particles. The impact of hydrodynamic interactions is reversed compared to force-driven motion, suggesting that our findings are a generic feature of flow-driven transport.

\end{abstract}

\maketitle

Collective transport of microscale particles in fluid media can occur either due to external forces or to a flow that drags the particles. Prominent examples 
include driven motion of particles through narrow channels~\cite{mirzaee-kakhki_colloidal_2020, lutz_singlefile_2004, wei_singlefile_2000}, 
in micro- and nanofluidic devices~\cite{Rodriguez-Villareal/etal:2020, petit_vesiclesonachip_2016, Ma/etal:2015}, in pores of zeolites \cite{VanDeVoorde/Sels:2017}, and through carbon nanotubes with relevance to biotechnological and biomedical applications \cite{Zeng/etal:2018}. 
Hydrodynamic interactions (HI) are always present in viscous fluids but only a few studies so far have tackled their influence on the collective transport behavior \cite{reichert_circling_2004, misiunas_densitydependent_2019, zahn_hydrodynamic_1997, Rinn/etal:1999, Malgaretti/etal:2012, Nagar/Roichman:2014, Grimm/Stark:2011}.
However, most of these studies have considered force- rather than flow-driven systems where HI were found to facilitate particle motion. 
In particular, it was reported that HI make it easier to surmount potential barriers when particles are force-driven across an optical sawtooth potential \cite{lutz_surmounting_2006}.
Here we show that the opposite is true for flow-driven transport of particles through periodic potentials. 
Using a combined experimental and theoretical approach we show that HI can lead to an effective enhancement of potential barriers, causing a slowing down of particle motion and a jamming at high density.

To experimentally realize a well controlled flow-driven system it is necessary to overcome several difficulties. 
For example, the pressure field should be precisely controllable, and one should be able to vary the \change{number} of particles and the height of the potential barriers.
Usually, flow-driven systems are realized in lithographically designed chambers, where the flow is produced by connecting a microfluidic channel to two particle reservoirs at different pressures.
However, for studying many-particle dynamics, such system could introduce unwanted effects, including fluctuations of the particle number in the channel and, even more importantly, the collective dynamics  may be affected by strong channel-reservoir coupling effects. 
These couplings can lead to nonequilibrium steady states, where the \change{particle number density} is determined by uncontrollable details \cite{Dierl/etal:2012}. 
We avoid such complications by experimentally generating a flow in a confined system with periodic boundary conditions. 
This is a typical setup considered in theoretical studies, which we realize here by driving particles dispersed in a liquid via a traveling-wave like potential along a ring. In a reference frame co-moving with the traveling wave, the particle transport then corresponds to that in a flow-driven system.

\begin{figure}[t]
\includegraphics{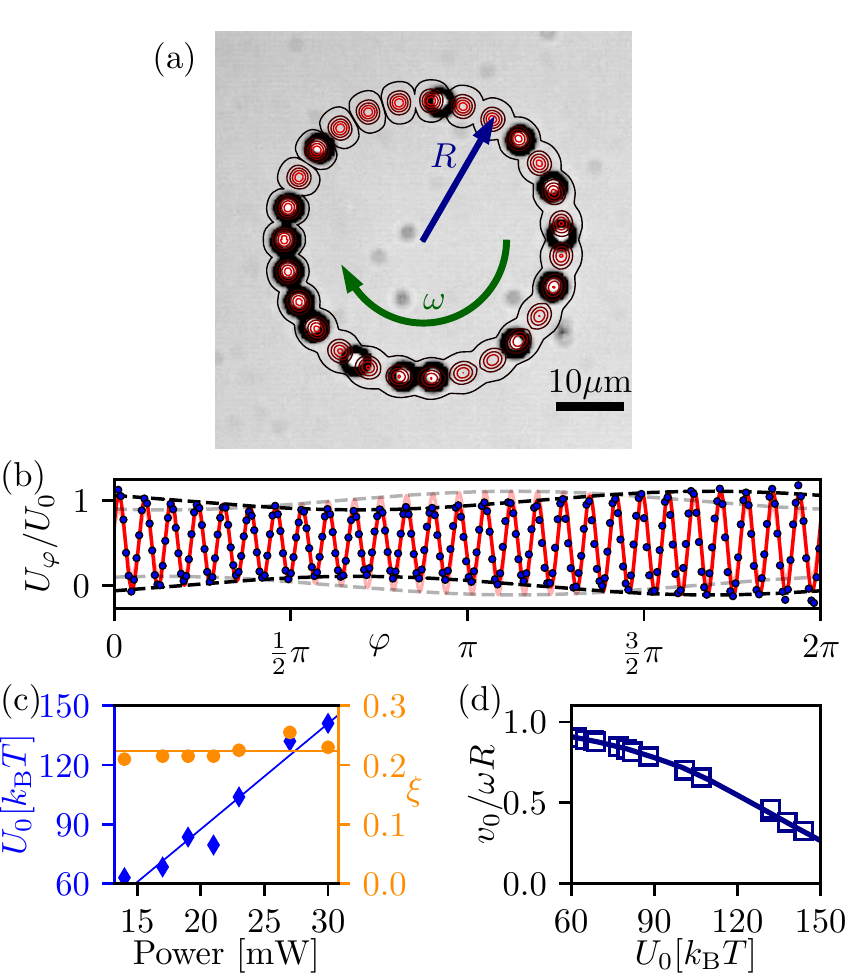}
\caption{
(a) Image of 15 particles (black circles) confined to a ring of radius $R$ with  27 equidistantly placed optical traps (contrast was enhanced by standard gamma correction). The ring rotates with angular velocity $\omega$. The lines mark isolines of the optical potential $U(r,\varphi)$. Along the azimuthal direction, the potential is given by $U_\varphi(\varphi,t_0)$ in 
Eq.~\eqref{eq:potential} for a fixed time $t_0$ and $r=R$.
(b) Potential $U_\varphi(\varphi, t_0)$ as a function of $\varphi$. The faded line shows the potential at a later instant $t_0+\Delta t$, and the dashed \change{black} lines indicate the potential amplitude modulation.
Experimental data (\change{markers}) are determined by averaging the position increments of a single particle measured at a sequence of times $t_n=t_0+2\pi n/\omega$, $n=0,1,2,...$ The \change{red} line is a fit of Eq.~\eqref{eq:potential} to these experimental data.
(c) Mean barrier height $U_0$ and modulation parameter $\xi$ as a function of the laser power.
(d) Average single-particle velocity $v_0$ normalized to the flow velocity $\omega R$ as a function of the mean barrier height $U_0$. Experimental values (symbols) are compared to simulations (line) for the potential in Eq.~\eqref{eq:potential}.}
\label{fig:exp_setup}
\end{figure}

Specifically, $N_{\rm tr}=27$~optical traps are positioned along a ring of radius 
$R=20.22\,\si{\mu m}$ in the $xy$-plane, see Fig.~\ref{fig:exp_setup}. We created the traps by passing an infrared 
continuous wave laser through two acousto-optic deflectors (AODs) that are capable to deflect the beam to a different position every $20\,\si{\mu s}$, thus each trap is generated once every $0.54\,\si{ms}$. 
\change{A fluid cell with an aqueous solution of spherical polystyrene colloids of radius $a=2\,\si{\mu m}$ (CML, Molecular Probes) is sealed by two coverslips separated by $\sim100\si{\mu m}$ and placed on a custom-built inverted optical microscope. The particles sediment by gravity close to the bottom cell and float there at a surface to surface distance of $\sim$200nm. Thus, they are far from the top plate and also from the in-plane boundaries located a few centimeters away.}
The microscope is equipped with a \change{CMOS camera (Ximea MQ003MG-CM) and tracking is done using the Crocker-Grier method \cite{crocker_methods_1996}. Note that the particle diameter
$2a=4\,\si{\mu m}$ is comparable to the distance $\lambda=2\pi R/N_{\rm tr}\cong4.7\,\si{\mu m}$ between neighboring traps.}


Due to the fast deflection of the laser beam, all traps appear as simultaneous because the typical self-diffusion time of the particles is $D_0/a^2 \sim 30$s, where $D_0\cong 0.1295\,\si{\mu m^2s^{-1}}$ is the particle diffusion coefficient. We independently obtained $D_0$ by measuring the time-dependent mean-square displacement in the absence of the optical ring.

Figure~\ref{fig:exp_setup}(a) illustrates the optical potential landscape $U(\bm{r})$, which confines the particles to a ring with potential valleys at the trap centers.
Expressed in polar coordinates, $U(r,\varphi)$ has a deep minimum along the radial direction at $r=R$.
Along the azimuthal direction it has $N_{\rm tr}$ pronounced minima at the trap positions shown by the dark red lines in Fig.~\ref{fig:exp_setup}(a). 
By changing the phase of the circulating beam linearly in time, the traps move at a constant angular velocity of \change{$\omega=0.63\,\si{rad\, s^{-1}}\cong 6\,\textrm{rpm}$} 
and the potential becomes time-dependent, $U=U(r,\varphi-\omega t)$. 
All of the following observations are done in the reference frame where the trap positions are 
stationary: $\varphi \rightarrow \varphi-\omega t$.

\begin{figure*}[t]
\centering
\includegraphics{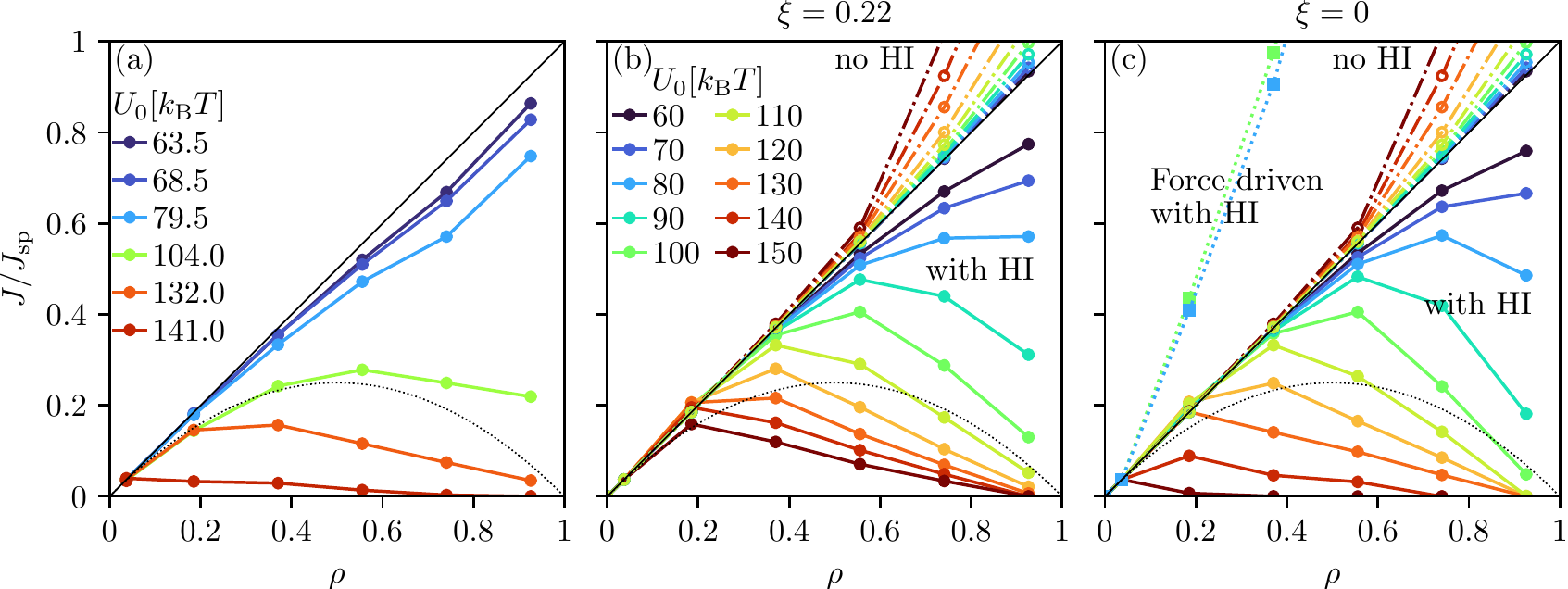}
\caption{Fundamental diagrams for (a) experiments and (b), (c) for simulations based on Eqs.~\eqref{eq:langevin}.
The current $J(\rho)$ is normalized with respect to the single-particle current 
$J_{\rm sp}=v_0/2\pi R$ where $v_0$ is the single-particle velocity from Fig.~\ref{fig:exp_setup}(d). Filled circles/solid lines and open circles/dashed-dotted lines refer to simulations with and without HI, respectively. 
In (b) results are shown for the time-dependent potential in Eq.~\eqref{eq:potential} with amplitude modulation ($\xi=0.22$). In (c) results are shown for the ideal periodic potential 
($\xi=0$), and for a force-driven system (squares/dotted lines). The legend in (b) applies also to (c). In all graphs
the solid black line marks the behavior for independent particles and the thin dotted line indicates the function $\rho(1-\rho)$ 
as a reference to jamming (see text).}
\label{fig:fund_diag}
\end{figure*}

From our setup depicted in Fig.~\ref{fig:exp_setup}(a), one should expect the potential to be periodic along the azimuthal direction with a period $2\pi/N_{\rm tr}$, corresponding to a wavelength $\lambda=2\pi R/N_{\rm tr}$. To check whether this is the case, and to obtain parameters for our simulations, we measured the optical forces by dragging a single particle along the ring \change{and analyzed particle displacements between successive frames}, similarly as in Ref. \cite{juniper_acoustooptically_2012}.
Videos S1 and S2 of the Supplemental Material~\cite{EPAPS} show that the particles exhibit negligible radial displacements, and thus we can consider the dynamics to proceed only along the ring through a one-dimensional potential $U_\varphi(\varphi)=U(R,\varphi)$. 
However, the potential turned out not to be perfectly $\lambda$-periodic, because the AOD response is not flat along the sample, causing the deflected laser power to vary weakly with $\varphi$.
\change{This leads to a static modulation of the potential amplitude with a relative strength $\xi$. In the corotating frame, 
the potential is periodically varying with angular frequency $\omega$, yielding a time-dependent potential}
\begin{equation}
U_\varphi(\varphi,t) = \frac{U_0}{2} [1+\xi \sin(\varphi+\omega t)]\cos\left(N_{\rm tr} \varphi\right)\,.
\label{eq:potential}
\end{equation}
The parameters of Eq.~(\ref{eq:potential}) were obtained by fitting the forces for various
laser powers in the range $11-40\,\si{m\!W}$. 
Figure~\ref{fig:exp_setup}(b) shows a typical outcome, where the dashed line represents the modulation \change{caused by the AOD’s response}.
As shown in Fig.~\ref{fig:exp_setup}(c), the modulation strength
$\xi = 0.22 \pm 0.02$ is almost independent of the laser power, while
the mean barrier height $U_0$ between the traps increases linearly with the laser power from 70 to 150$\,k_{\rm B}T$, where $k_{\rm B}T$ is the thermal energy.

To check the parameters $U_0$, $\xi$, and $D_0$, we performed
Brownian dynamics simulations (see below) and compared simulated with 
measured drag velocities $v_0$ for the amplitudes $U_0$ applied in the experiment. The results in Fig.~\ref{fig:exp_setup}(d) show an excellent agreement between simulated and measured data.

We now study the fundamental diagram \cite{Schadschneider/etal:2010} of the many-particle dynamics, i.e.\
the relation between the current $J$ along the azimuthal direction and the particle density $\rho$. 
To determine the current, we first calculated the instantaneous velocities of each particle in the stationary state \change{from changes of the particle positions in small time steps (successive frames)}
along the azimuthal direction. 
We then averaged these velocities over time and all particles.
The resulting mean velocity $\langle v\rangle$ gives the current $J=N\langle v\rangle/(2\pi R)$. 
For the density, we use the fraction of traps occupied by the particles, $\rho=N/N_{\rm tr}$. 

The current $J(\rho)$ is shown in Fig.~\ref{fig:fund_diag}(a) for various barrier heights $U_0$. To see the impact of the particle interactions, we normalized $J(\rho)$ with respect
to the current of a single-particle, $J_{\rm sp}=v_0/(2\pi R)$. 
For independent particles, we would expect a linear function $J(\rho)/J_{\rm sp}=\rho$, represented by the straight black line in the figure. 
All curves $J(\rho)/J_{\rm sp}$ approach this line in the limit $\rho\to0$, while
for $\rho>0$ the current is smaller, and as $U_0$ is increased, the current suppression is larger. 

When $U_0\lesssim 80 k_{\rm B}T$, the suppression of the current is 
relatively weak for all $\rho$, while at larger $U_0\gtrsim 100 k_{\rm B}T$, we find a strong suppression at large $\rho$. 
A local maximum in $J(\rho)$ occurs, which is a signature of jamming, i.e.\ a transition from a fluid-like 
continuous motion to a thermally activated behavior with blocking effects.  
A prominent simple model for this jamming is the asymmetric simple exclusion process \cite{Derrida:1998, Schuetz:2001} where the current follows,
$J(\rho)/J_{\rm sp}=\rho(1-\rho)$, which we included in 
Fig.~\ref{fig:fund_diag} as a reference (black dotted line). However, the dynamics in our experiment are more complex as witnessed
by the more complicated shapes of the current-density relations and their strong dependence on $U_0$.

The jamming-like behavior seen in Fig.~\ref{fig:fund_diag}(a) suggests that it arises from the strong confinement of the particles along the ring,  which hinders them from overtaking each other (single-file transport). Is this jamming originating solely from a hard-sphere type interaction between the  polystyrene colloids? Or is the combined effect of this interaction with the time-dependence of the potential in Eq.~\eqref{eq:potential} essential?  And what is the impact of HI?

To answer these questions, we performed Brownian dynamics simulations of hard-sphere interacting particles.
The optical forces acting on the particles at positions $\bm r_i$ are described by the potential $U(\bm{r}_i-\bm u_i t)$, where
$\bm u_i=\bm{\omega}\times\bm r_i$ are the azimuthal velocities of the particles 
($\bm\omega=\omega\hat{\bm z}$, where $\hat{\bm z}$ is the unit vector in $z$ direction).

In our reference frame, which corotates with the traps, the translational motion of $N$ particles in the presence of HI
can be described by the Langevin equations \cite{Ermak/McCammon:1978}
\begin{subequations}
\label{eq:langevin}
\begin{eqnarray}
\dot{\bm{r}_i}&=&-\bm\omega\times\bm r_i + \sum_{j=1}^{N}
\bigl[k_{\rm B}T\bm\nabla_j \bm\mu_{ij}+\bm\mu_{ij}\bm f_i\bigr] + \bm\eta_i,\label{eq:langevin-a}\\
\bm f_i&=&-\bm\nabla_i U(\bm{r}_i)+\bm f_i^{\rm int}\,,\label{eq:langevin-b}
\end{eqnarray}
\end{subequations}
where $\bm f_i^{\rm int}$ is the interaction force exerted on particle $i$ by the other particles, and
$\bm{\mu}_{ij}=\bm{\mu}_{ij}(\bm r_1,\ldots,\bm r_N)$ is the mobility tensor, which accounts for HI by the mobility method 
\cite{Kim/Karrila:1991}. 
The vector $\bm\eta_i$ is a Gaussian white noise 
with zero mean and covariance matrix in accordance with the fluctuation-dissipation theorem, i.e.\
$\langle\bm\eta_i(t)\otimes\bm\eta_j(t')\rangle=2k_{\rm B} T\bm\mu_{ij}\delta(t-t')$. 

Equation~\eqref{eq:langevin-a} shows that our experiment indeed corresponds to a flow-driven system of 
interacting particles. The term $\bm\omega\times\bm r_i$ equals the driving in a radially symmetric vortex flow 
field. \change{When averaging $U_\varphi(\varphi,t)$ from Eq.~\eqref{eq:potential} over one time period $2\pi/\omega$, it becomes time-independent and 
periodic in $\varphi$ with period $2\pi/N_{\tr}$. Accordingly, when averaging the force of the time-averaged 
potential over one period $2\pi/N_{\tr}$ along the azimuthal direction, it is zero. Shortly speaking,} in a fluid at rest, there 
would be no particle current on average \footnote{For asymmetric potentials as those considered in Brownian motors, there can be a net driving also if forces are zero after period-averaging but this does not apply here because of lack of asymmetry.}.

To take the HI into account, we used the procedure developed in 
\cite{Hansen/etal:2011}, where the mobilities $\bm{\mu}_{ij}$
are given by the Rotne-Prager form, both for the reflective
fluid flows resulting from the coverslip underneath the particles and the fluid flows induced by the movements of the 
particles.
This consideration of HI constitutes a minimal model, 
where we neglect lubrication effects, possible translation-rotation couplings and expansions of the mobilities beyond the 
Rotne-Prager form for particles coming close to each other. 
In all simulations we included the reflective fluid flow and constrained the particle movement to the azimuthal direction. When we refer to  simulations with and without HI in the following, we mean the particle-particle HI, which we calculated in full three-dimensional space.

Figure~\ref{fig:fund_diag}(b) shows the simulated normalized current $J(\rho)/J_{\rm sp}$ obtained for the potential~\eqref{eq:potential}
with amplitude modulation as in the experiment ($\xi=0.22$). 
The open circles connected by the dashed-dotted lines refer to the simulations without HI. 
In contrast to a suppression of the current, they show an enhancement at larger densities. 
\change{This enhancement is due to the fact that with increasing density, regions of large driving
force are more strongly populated.}
The behavior changes drastically when including the HI (filled circles connected by solid lines). 
The normalized current now decreases at 
large $\rho$ in qualitative agreement with the experiments. 
In particular, the jamming-like behavior becomes stronger with increasing $U_0$. The details of the behavior differ from the experiments in Fig.~\ref{fig:fund_diag}(a), and we attribute this to 
the limitations of our minimal model of HI.

To clarify whether the amplitude modulation plays a crucial role for the jamming, we performed additional simulations
for the ideal case when the periodic potential is time-independent ($\xi=0$ in Eq.~\eqref{eq:potential}). 
The results shown in Fig.~\ref{fig:fund_diag}(c) are very similar to those in Fig~\ref{fig:fund_diag}(b).
Without HI, $J(\rho)$ exceeds the single-particle current. In the presence of HI, the current is smaller than that of independent particles and
decreases with $\rho$ at large densities, reflecting jamming. 

In order to explore if the jamming would occur in a force-driven system, we 
solved the Langevin equations~\eqref{eq:langevin} for particles driven by the external torque 
$\bm f_{\rm ext}(\bm\hat{\bm\varphi})=-\omega R\hat{\bm\varphi}$, where $\hat{\bm\varphi}$ denotes the unit vector along the azimuthal direction; 
\change{this means that Eq.~\eqref{eq:langevin-a} becomes
$\dot{\bm{r}_i}=\sum_j
\left\{k_{\rm B}T\bm\nabla_j \bm\mu_{ij}+\bm\mu_{ij}\left[\bm f_i+\bm f_{\rm ext}(\hat{\bm\varphi}_i)\right]\right\}+ \bm\eta_i$}. In contrast to the flow-driven case, the results
for the force-driven one show a strong current enhancement
in agreement with previous findings \cite{reichert_circling_2004}, see the squares connected by dotted lines in Fig.~\ref{fig:fund_diag}(c).

We now provide an argument for the current suppression in flow-driven transport. 
To this end, we consider the equation of motion of a single particle in the potential given by Eq.~\eqref{eq:potential},
\begin{equation}
\dot\varphi
=-\frac{\mu_0}{R^2}\frac{\partial}{\partial\varphi}\left[
\frac{\omega R^2}{\mu_0}\varphi+U_\varphi(\varphi,t)\right]+\eta_\varphi\,,
\end{equation}
where $\langle \eta_\varphi(t)\eta_\varphi(t')\rangle=2D_0R^{-2}\delta(t-t')$ with $D_0=k_{\rm B}T\mu_0$. 
Hence, $U_\varphi(\varphi,t)$ is tilted by the linear potential $\omega R^2\varphi/\mu_0$, resulting in the
effective potential $U_\varphi^{\rm eff}(\varphi,t)=U_\varphi(\varphi,t)+\omega R^2\varphi/\mu_0$.
For small $U_0$, this effective potential exhibits no barriers, implying that a single particle is essentially dragged by the flow.
The part $U_\varphi$ in $U_\varphi^{\rm eff}$ can be viewed as creating a resistance for the flow-driven particle motion, causing
$v_0$ to be smaller than the flow velocity $\omega R$ for $U_0>0$ and to decrease with $U_0$.

When $U_0$ becomes larger than a critical amplitude $U_{0{\rm c}}$,
the single-particle motion becomes thermally activated. For determining $U_{0{\rm c}}$, we take
the temporally period-averaged potential
$\bar U_\varphi^{\rm eff}(\varphi)=(\omega/2\pi)\int_0^{2\pi/\omega}U_\varphi(\varphi,t)\dd t$, which
is also the effective potential for the ideal case $\xi=0$ [time-independent periodic potential, see Eq.~\eqref{eq:potential}].
The barriers in $\bar U_\varphi^{\rm eff}(\varphi)$
emerge when $U_0$ passes the critical amplitude
$U_{0{\rm c}}=2\omega R^2/\left(\mu_0N_{\rm tr}\right)\cong 147\,k_{\rm B}T$.

The above shows that all experiments are below the critical amplitude, and therefore one would expect linear current-density relations with no pronounced jamming effect. 
This is indeed confirmed by the simulated data without HI shown in Fig.~\ref{fig:fund_diag}(b). 
However, the HI leads to an effective increase of the potential barrier. 
Within the Rotne-Prager level of description, the terms $\bm\mu_{ij}\bm f_j$ contain the part
$(\bm\mu_{ij})_{\varphi\varphi}\partial U_\varphi/\partial\varphi$, where $(\bm\mu_{ij})_{\varphi\varphi}$ is the
$\varphi\varphi$-component of the tensor $\bm\mu_{ij}$ in cylindrical coordinates. In the absence of HI, only
the term $\mu_0\partial U_\varphi/\partial\varphi$ for $j=i$ contributes to the sum over $j$.  In the presence of HI, the additional contributions for $j\ne i$ are dominated by the particles $j$ closest to particle $i$. For a particle $j$ located at a neighboring trap of particle $i$, in particular, the corresponding additional contribution is 
$\mu_0[(3a/2\lambda)-(a/\lambda)^3]\partial U_\varphi/\partial\varphi$
\change{($a$: particle size, $\lambda$: wavelength of the potential).}

This can be viewed as an effective enhancement of the mean barrier height $U_0$ by HI to a value
\begin{equation}
U_0^{\rms HI}=\left(1 + \frac{3a}{2\lambda} -  \frac{a^3}{\lambda^3} \right) U_0\,.
\label{eq:U0HI}
\end{equation}
Because $a<\lambda$, the additional contribution is always positive, i.e.\ $U_0^{\rms HI}>U_0$. Using Eq.~\eqref{eq:U0HI} with  
$\lambda=2\pi R/N_{\rm tr}\cong4.71\,\si{\mu m}$ and $a\cong 2\,\si{\mu m}$, 
$U_0$ is increased by about 56\%. 
For large $U_0$ in experiment, this gives $U_0^{\rm HI}>U_{0{\rm c}}$, which implies that the effective potential 
$U_\varphi^{\rm eff}$ (now for one particle in the many-particle system) exhibits barriers.
The dynamics thus becomes thermally activated and strongly slows down. 
This barrier enhancement effect should become stronger
with larger occupation probabilities of neighboring traps, i.e.\ density. In addition, blocking effects can lead to a further slowing down with increasing $\rho$ \cite{Lips/etal:2018, Lips/etal:2019}.
Indeed, we have seen indications of a hopping motion and jamming
in the measured trajectories for large $U_0$ and high densities, see videos in the Supplemental Material \cite{EPAPS}.

Our analysis can be made more precise by refining the description beyond the Rotne-Prager approximation.
This can include higher order terms in the expansion of mobilities in powers of the particle radius to interparticle distance ratio 
$a/\lambda$, lubrication effects, and an additional consideration of the rotational dynamics of the particles. 
One can imagine to reach a quantitative agreement between experiments and modeling
with large-scale simulation methods, as, for example, by implementing multiparticle collision dynamics \cite{Howard/etal:2019}.
However, our model is able to qualitatively reproduce the experimental findings and it provides an understanding of the mechanisms governing the current suppression in our system. 

To conclude, we have found that HI in flow-driven many-particle systems can lead to an effective barrier enhancement, 
which induces a jamming-like behavior. We support this conclusion by combining experiments with numerical simulations and we provide a minimal approach for exploring the impact of HI on the translational motion. 
The flow-driving is essential for the observed phenomena and the reported effect is present regardless of imperfections of the external potential, which makes our findings relevant for non-ideal situations in nature.

Particle transport over energetic barriers occurs in many soft matter and biological systems. We expect that the phenomena uncovered here, namely the
HI-induced barrier enhancement and jamming effects will be of general importance for further studies
and applications of flow-driven many-particle systems.

\acknowledgements{We sincerely thank G.\ N\"agele, H.\ Stark for advice, and the members of the DFG Research Unit FOR 2692 for fruitful discussions. 
E. C.-L., A. O.-A and P. T., acknowledge support from the ERC Consolidator Grant (Grant agreement number 811234). 
P. T.  acknowledges support from Ministerio de Ciencia, Innovación y Universidades (PID2019-108842GB-C21), AGAUR (2017-SGR-1061) and Generalitat de Catalunya under Program ``ICREA Acad\`{e}mia''.
A.R., P.M., and D.L.\ gratefully acknowledge financial support by the Czech Science Foundation (Project No.\ 20-24748J) and the Deutsche Forschungsgemeinschaft (Project No.\ 432123484).}

\bibliography{bibliography.bib}
\end{document}